\documentclass[aps,prd,reprint,groupedaddress,showpacs,nofootinbib,amsmath]{revtex4-1}

\usepackage{amsmath}
\usepackage{amssymb}
\usepackage{physics}
\usepackage{bbold}
\usepackage{graphicx}
\usepackage{verbatim}
\usepackage{hyperref}
\usepackage{indentfirst}
\usepackage{mathtools}
\usepackage{xcolor}
\usepackage{color}
\usepackage{adjustbox}
\usepackage{placeins}
\usepackage{lipsum}
\usepackage{csquotes}
\usepackage{hyperref}

\begin{document}

\title{Baryogenesis in cosmological models with symmetric and asymmetric quantum bounces}


\author{P.~C.~M.~Delgado}\email{pmordelgado@gmail.com}
\affiliation{CBPF - Centro Brasileiro de
Pesquisas F\'{\i}sicas, Xavier Sigaud st. 150,
zip 22290-180, Rio de Janeiro, Brazil.\\
PPGCosmo, CCE - Universidade Federal do Espírito Santo, Avenida Fernando Ferrari 514, zip 29075-910, Vitória, Brazil.}

\author{M.~B.~Jesus}\email{marcusbomfimdejesus@yahoo.com.br}
\affiliation{CBPF - Centro Brasileiro de
	Pesquisas F\'{\i}sicas, Xavier Sigaud st. 150,
	zip 22290-180, Rio de Janeiro, Brazil.}

\author{N.~Pinto-Neto}\email{nelson.pinto@pq.cnpq.br}
\affiliation{CBPF - Centro Brasileiro de
	Pesquisas F\'{\i}sicas, Xavier Sigaud st. 150,
	zip 22290-180, Rio de Janeiro, Brazil.}

\author{T.~Mourão}\email{tiagomcsphysics@gmail.com}
\affiliation{CBPF - Centro Brasileiro de
	Pesquisas F\'{\i}sicas, Xavier Sigaud st. 150,
	zip 22290-180, Rio de Janeiro, Brazil.}

\author{G.~S.~Vicente}\email{gustavo@fat.uerj.br}
\affiliation{FAT - Faculdade de Tecnologia, UERJ - Universidade do Estado do Rio de Janeiro, Rod. Pres. Dutra, km 298,
	zip 27537-000, Resende, Brazil.}

\date{\today}

\graphicspath{ {./graphics/} }

\begin{abstract}
The baryon$-$antibaryon asymmetry (excess of matter over antimatter in our Universe), indicated by 
observational data from the Cosmic Microwave Background anisotropies, 
predictions of primordial Nucleosynthesis, and 
the absence of intense radiation from matter-antimatter annihilation, 
constitutes an unsolved puzzle in cosmology. Two mechanisms for baryon asymmetry have been proposed as extensions of the Standard Model of Particle Physics
at high energies. They rely on new couplings involving the baryon number current, one with a scalar field, called Spontaneous Baryogenesis, and the other
with space-time curvature, named Gravitational Baryogenesis. These two mechanisms are investigated in the context of many bouncing scenarios, either symmetric or asymmetric around the bounce. It is shown that the constraints on the free parameters of these scenarios, imposed to yield the observed baryon-to-entropy ratio, 
are mild, already containing the values compatible with other observational constraints coming from the features of the power spectra of cosmological perturbations.
Hence, realistic bouncing models can yield the observed baryon$-$antibaryon asymmetry if one of the two mechanisms proposed takes place in nature.\\

$\copyright$ 2020 American Physical Society \
DOI: \href{https://journals.aps.org/prd/abstract/10.1103/PhysRevD.102.063529}{10.1103/PhysRevD.102.063529} 
\end{abstract}


\maketitle

\section{Introduction}

The standard cosmological model is based in General Relativity (GR), and its initial singularity represents a major problem.
However, this problem can be overcome invoking quantum mechanics, where the singularity
can be avoided by the occurrence of quantum effects beyond classical GR, which become important at small scales.
Indeed, non-singular universes emerge in this framework, under many approaches \cite{pbb,cai,Pinto-Neto:2013toa,lqc,ekp}, and the hot Big-Bang initial singularity is eliminated. 
If expansion does not begin at the singularity, there must be a preceding phase, which can be a contracting phase, 
in this case yielding a bouncing universe. 

In the framework of quantum cosmology in minisuperspace models, non-singular bounces occur due to quantum 
effects in the background~\cite{n6,n7,n8,PintoNeto:2009gv}. 
The standard Copenhagen interpretation of quantum mechanics, however, does not apply to the universe as a whole 
due to the fact that it demands an external agent in order to perform a measurement, and establish a physical reality from the collapse of a probability amplitude.
The de Broglie-Bohm theory~\cite{dBB} is, among others~\cite{n10,np2}, 
independent of an external agent. In this theory, a collapse assumption is no longer necessary, so that the universe is a deterministic 
system and real (Bohmian) trajectories exist.  
This interpretation can be applied to cosmology, where the universe is described by the quantum evolution of the scale factor, $a(t)$, 
derived from a wave function solution of a quantum cosmology equation, in many approaches being the Wheeler-DeWitt equation~\cite{wdw}. 

The de Broglie-Bohm (dBB) quantum bounce contains a contracting phase, which last until the scale factor reaches a minimum value,
followed by a phase of expansion, which we recognize as the usual Friedmann expansion phase, when the classical limit is recovered. 
The physics around the bounce is governed
by quantum effects, whereas the phases of contraction and expansion are described by the standard cosmology.

The evolution from contracting to expanding phase can be symmetric~\cite{Pinto-Neto:2013toa,Peter:2008qz}.
An example is the case where the matter contents are radiation and dust~\cite{PintoNeto:2005gx}, described by perfect fluids, 
which has the following evolution: dust contracting phase, radiation contracting phase, bounce dominated by radiation, 
radiation expanding phase and dust expanding phase. The presence of the dust fluid, which can be dark matter, is important in 
order to obtain an almost scale invariant spectrum of scalar cosmological perturbations. 

A more involved bounce dynamics is given in Ref~\cite{Bacalhau:2017hja}, where a single scalar field with exponential potential drives the bounce as a stiff matter fluid, behaves as a dust fluid in the asymptotic past and future, and also presents a transient dark energy-type behavior occurring only in the future of the expanding phase. This bounce is asymmetric because the transient dark energy epoch occurs only in the expanding phase, not in the contracting phase, avoiding problems 
related to the imposition of vacuum state initial conditions in the contracting phase if dark energy is present there, and overproduction of gravitational waves, which
are typical in bouncing models containing a canonical scalar field. 
Other asymmetric bounces where obtained in Ref.~\cite{Delgado:2020htr}, with either unitary and non-unitary evolution\footnote{In dBB quantum cosmology, it is not necessary to impose a probabilistic interpretation to the wave function of the whole system. Only for the so called conditional wave functions, which apply to sub-systems of the whole system, and satisfy an effective Schr\"odinger equation with unitary evolution, does a probability notion emerges. See Ref.~\cite{Falciano:2008nk,ward} for details.}.
One particular interesting result was one solution describing an expanding cosmological universe arising from an almost flat space-time. 
This type of asymmetry is particularly relevant because these solutions may be used to account for non-negligible back-reaction due to quantum 
particle production around the bounce (see Refs.~\cite{Celani:2016cwm,Scardua:2018omf}), which is important for the study of baryogenesis.

The baryon$-$antibaryon asymmetry (excess of matter over antimatter in our Universe) indicated by 
observational data from Cosmic Microwave Background~\cite{Agashe:2014kda}, 
predictions of Big-Bang Nucleosynthesis~\cite{Aghanim:2018eyx}, and 
the absence of intense radiation from matter-antimatter annihilation~\cite{Cohen:1997ac} 
constitutes an unsolved puzzle in cosmology. 
The baryon-to-entropy ratio is $n_B/s = 9.2^{+0.6}_{-0.4}\times10^{-11}$.
The current view is based on the Sakharov conditions~\cite{Sakharov:1967dj}, which should hold during the early hot Universe, yielding a net baryon asymmetry, and cease to be satisfied as the Universe expands and cools.
An important theory on this subject is electroweak baryogenesis~\cite{Kuzmin:1985mm,Shaposhnikov:1986jp,Shaposhnikov:1987tw}, which satisfies all Sakharov conditions. However, it is unable to yield sufficient baryon asymmetry within the Standard Model of Particle Physics (SMPP). In order to solve this issue one needs to explore physics beyond the SMPP, like in Ref.~\cite{bediaga}.
Another relevant mechanism is the so called {\it Spontaneous Baryogenesis}~\cite{Cohen:1987vi,DeSimone:2016ofp}, which is based in the coupling of a scalar field to the baryon number current.
The main point is that baryon asymmetry is generated while baryon violating interactions are still in thermal equilibrium, which is not in contradiction with the Sakharov conditions because the scalar field coupling in a expanding universe violates CPT invariance.
A third mechanism, which is termed {\it Gravitational Baryogenesis}~\cite{Davoudiasl:2004gf}, is a natural extension of the latter mechanism, which can naturally occur in an effective theory of gravity. Introducing a coupling between the derivative of the Ricci scalar and the baryon number current, this interaction gives opposite signs for energy contributions to particles and antiparticles, also violating CPT symmetry. This induces changes in the thermal equilibrium distributions which result in a nonzero net baryon number.
These two mechanisms are similar: the new interaction terms violate CP and is CPT conserving {\it in vacuo}, but both dynamically break CPT in an expanding universe, where, in the Gravitational Baryogenesis case, the curvature varies in time, or, in the Spontaneous Baryogenesis case, where the scalar field, not being in its vacuum state, drives the cosmological evolution, and hence evolves in time.
It is important to point out that a bounce solution is important for this mechanism to be effective. From Ref.~\cite{Odintsov:2016apy} in the context of Loop Quantum Cosmology, one can notice from its Eq.~(14) that a radiation dominated universe cannot produce baryon asymmetry in the Einstein-Hilbert case (critical density $\rho_c\to\infty$), whereas for a bouncing universe (finite $\rho_c$), it becomes possible. 

In this paper we present the baryogenesis scenario in the context of dBB quantum cosmology, in both symmetric and asymmetric non-unitary and unitary realizations, based on the results of Refs.~\cite{Delgado:2020htr,Bacalhau:2017hja}. 
We consider both Spontaneous and Gravitational Baryogenesis mechanisms, which we call {\it Baryogenesis with Scalar Coupling} and {\it Baryogenesis with Curvature Coupling}, respectively.

The paper is outlined as follows.
In Sec.~\ref{secII}, we present the main aspects of standard cosmological baryogenesis, which are subjected to the Sakharov conditions.
The mechanisms of spontaneous and gravitational baryogenesis are introduced, stressing the fact that the third Sakharov condition can be overcome in these frameworks.
We work these conditions in detail for a hypothetical decay, and we show that baryon asymmetry can take place in thermal equilibrium.
In Sec.~\ref{secIII}, we introduce the background bouncing models, i.e., the mini-superspace models in 
the dBB theory. Firstly, the standard symmetric quantum bouncing trajectories are obtained from initial static Gaussian wave functions centered at the origin, i.e., without phase velocity. Secondly, we present asymmetric quantum bounce trajectories for a non-unitary wave function from an initial Gaussian wave function with nonzero phase velocity and for an unitary wave function from a superposition of Gaussian wave functions multiplied by factors of the form $\operatorname{exp}(ip^2\chi^2)$. These new quantum parameters are responsible for the asymmetry.  
In Secs.~\ref{secIV} and~\ref{secV}, we analyse the gravitational and spontaneous baryogenesis mechanisms for the models presented in Sec.~\ref{secIII}, and the one
presented in Ref~\cite{Bacalhau:2017hja}. Some analytical results are obtained for the baryon-to-entropy ratio, and constraints on the physical parameters of the theory are obtained.
In the Conclusion, section~\ref{conclusions}, we summarize and comment the results, and discuss future perspectives.

\

\section{Cosmological Baryogenesis}
\label{secII}

Proposals of baryogenesis mechanisms are traditionally concerned with satisfying the three Sakharov's conditions~\cite{Sakharov:1967dj},
which are: A) violation of the baryon number $B$, B) violation of $C$ and $ CP $ and C) thermal equilibrium deviation. 

The understanding of these conditions is easily perceived by analyzing a hypothetical decay. Suppose that a particle $ X $ decays only in two channels, which produces the baryon numbers $ B_1 $ and $ B_2 $, where the respective decay rates are $ \Gamma (X \rightarrow q_1q_1) $ and $ \Gamma (X \rightarrow q_2q_2) $. Then, the $X$ total decay rate is of the form:
\begin{equation}
    \Gamma_X=\Gamma(X\rightarrow q_1q_1) + \Gamma(X\rightarrow q_2q_2).
\end{equation}
Therefore, the probability that $ X $ will decay on the channel producing the number $ B_1 $ is given by:
\begin{equation}
    r=\Gamma(X\rightarrow q_1q_1)/\Gamma_X,
\end{equation}
where the channel associated with baryon number $ B_2 $ has a complementary probability of occurrence, i.e., $ 1-r $.

The decay of the antiparticle of $ X $, $ \bar{X} $, yields the baryonic numbers $ \bar{B}_1 = -B_1 $ and $ \bar{B}_2 = -B_2 $, which, in turn, have probabilities respectively given by $ \bar{r} $ and $ \bar{r} -1 $. Therefore, the baryon number difference is simply
\begin{equation}\nonumber\label{eqn:variação bar}
\Delta B = rB_1 + (1-r)B_2 - \bar{r}\bar{B}_1 - (1-\bar{r})\bar{B}_2
\end{equation}
\begin{equation}\label{eqn: variação de baryons}
= (r-\bar{r})(B_1-B_2).
\end{equation}
The total baryon number variation requires that $ B_1 $ and $ B_2 $ to be different baryon numbers, hence, the necessity of condition A. However, the decay probabilities of particles and their antiparticles should also be different, leading to condition B, as we will now see.

The $CPT$ symmetry imposes that the total particle decay rates and their associated antiparticles to be equal: $\Gamma_X = \Gamma_{\bar{X}}$.
When we inspect a simple decay channel, such as $\Gamma(X\rightarrow qq)$,  the necessity of $C$ violation in this context becomes clear. Indeed, 
\begin{eqnarray}
\label{eqn:pro c 1}
r &=&\frac{\Gamma(X\rightarrow qq)}{\Gamma_X},\\
\label{eqn:pro c 2}
\bar{r}&=&\frac{\Gamma(\bar{X}\rightarrow \bar{q}\bar{q})}{\Gamma_X}.
\end{eqnarray}
Therefore, if charge conjugation is a valid symmetry,
\begin{equation}
 r=Cr=\bar{r} \Rightarrow r-\bar{r} =0,
\end{equation}
hence $C$ must be violated.
Suppose now that $CP$ symmetry, which consists of charge and parity transformations, is valid, even if $C$ is violated. 
Hence, for the hypothetical probability decay channel,
\begin{eqnarray}
    r &=&\frac{\Gamma(X\rightarrow q_Lq_L)+\Gamma(X\rightarrow q_Rq_R)}{\Gamma_X},\\ \bar{r}&=&\frac{\Gamma(\bar{X}\rightarrow \bar{q}_L\bar{q}_L)+\Gamma(\bar{X}\rightarrow \bar{q}_R\bar{q}_R)}{\Gamma_X},
\end{eqnarray}
the $CP$ transformation implies that
\begin{eqnarray}
    CP[\Gamma(X\rightarrow q_Lq_L)]&=& \Gamma(\bar{X}\rightarrow \bar{q}_R\bar{q}_R), \\
CP[\Gamma({X}\rightarrow q_Rq_R)]&=& \Gamma(\bar{X}\rightarrow \bar{q}_L\bar{q}_L).
\end{eqnarray}
If we assume that $ CP $ is a symmetry of this decay, we get
\begin{eqnarray}
    CP[\Gamma(X\rightarrow q_Lq_L)]&=&\Gamma(X\rightarrow q_Lq_L),\\
CP[\Gamma({X}\rightarrow q_Rq_R)]&=& \Gamma({X}\rightarrow q_Rq_R),
\end{eqnarray}
so it is clear that
\begin{eqnarray}
r-\bar{r} &=&
\frac{\Gamma({X}\rightarrow q_L q_L)+\Gamma(X\rightarrow q_R q_R)}{\Gamma_X}\nonumber\\
    &-&\frac{\Gamma(\bar{X}\rightarrow \bar{q}_L\bar{q}_L)+\Gamma(\bar{X}\rightarrow \bar{q}_R\bar{q}_R)}{\Gamma_X}\nonumber\\
    &=&0,
\end{eqnarray}
leading that the difference in the amount of baryons also to be null in~Eq.~\eqref{eqn: variação de baryons}.

Finally, to explain the  third Sakharov's criterion, it is enough to calculate the average of the baryon number operator $\hat B$ in thermal equilibrium at a temperature $T = 1 / \beta$, which reads:
\begin{align}
    \langle \hat B \rangle_T &=Tr\left[ e^{-\beta H}\hat B \right] \nonumber\\
                         & =Tr\left[ (CPT)(CPT)^{-1}e^{-\beta H}\hat B \right]\nonumber\\
                         & =Tr\left[ e^{-\beta H}(CPT)^{-1}\hat B(CPT)  \right]\nonumber\\
                         & = - Tr\left[ e^{-\beta H}\hat B \right]\nonumber\\   
                         & = - \langle \hat B \rangle_T.
\end{align}
We have considered the fact that the Hamiltonian $ H $ commutes with the $ CPT $ operator. Thus, in thermal equilibrium, there is no mean baryon generation, i.e., $\langle \hat B \rangle_T = 0$.

Concluding, the usual approaches to baryogenesis rely on finding situations in the Universe where the three Sakharov conditions are satisfied. The spontaneous and gravitational baryogenesis scenarios \cite{Cohen:1987vi,DeSimone:2016ofp,Davoudiasl:2004gf}, however, take another route. The new couplings with the baryon current they propose, either with a scalar field or the curvature of space-time, lead
to a violation of $CPT$ invariance in a time dependent spacetime, as the Friedmann model. Hence, baryon generation takes place in thermal equilibrium, given that the Sakharov first condition holds. 

In this case, in high temperatures, where we expect the mechanisms which do not conserve baryon number take place, the difference between the number density of 
baryons and antibaryons reads (see, e.g. Ref.~\cite{Mukhanov})
\begin{equation}\label{eqn:n0}
n_B-n_{\bar B}=\frac{g_B T^3}{3}\frac{\mu_B}{T},
\end{equation}
where $g_B$ is the number of degrees of freedom of baryons, and $\mu_B$ is the chemical potential associated with the baryon number density through the new 
coupling with the baryon current proposed in these two scenarios.
The total entropy density, in turn, is given by \cite{Mukhanov},
\begin{equation}\label{eqn: entropia}
    s=\frac{4 {\pi}^2g_{*} T^3}{45},
\end{equation}
where $g_ {*}$ is the total multiplicity of relativistic degrees of freedom, containing all degrees of freedom of bosons, $ g_b $, 
and fermions, $ g_f $, which has the form:
\begin{equation}
    g_{*}=\sum_i g_{b,i} + \frac{7}{8}\sum_j g_{f,j}.
\end{equation}
The baryon-to-entropy ratio is then given by
\begin{equation}\label{eqn: razão BF}
       \frac{n_B}{s}= \frac{15 g_B \mu_B}{4\pi^{2}g_{*}T}.
\end{equation}

The mechanisms of gravitational and spontaneous baryogenesis will be discussed in Sections IV and V, separately. They lead to different 
$\mu_B$, but the whole calculation also depends on the particular time evolution of the background Friedmann model which is being considered. In the next
section, we describe the models we will investigate.

\section{The Background Bouncing Models}\label{secIII}

We consider cosmological models that arise through the Wheeler-DeWitt quantization of the background, taking into account the dBB interpretation of Quantum Mechanics \cite{dBB}. The latter solves the measurement problem through an effective collapse of the wave function, which is a consequence of the deterministic character of this interpretation. Thus, an external classical domain is not required in order to describe the measurement process, and we are able to quantize the entire universe \cite{Pinto-Neto:2013toa,ward}. The procedure of quantization and the bounce solutions are detailed in \cite{Delgado:2020htr}, and we will only mention the most relevant results for the present work. The other bounce solution given in Ref.~\cite{Bacalhau:2017hja} will be discussed in section V.

The Wheeler-DeWitt equation to be satisfied for a flat, homogeneous and isotropic universe filled with a perfect fluid with equation of state $P=\omega\rho$, where $P$ is the pressure, $\rho$ the energy density and $\omega$ the equation of state parameter, for a particular choice of ordering in $a$ (in which the Wheeler-DeWitt equation is covariant under redefinitions of the scale factor) reads
\begin{equation}\label{wdw equation}
i \frac{\partial \Psi(\chi,T)}{\partial T}=\frac{1}{4} \frac{\partial^{2}\Psi(\chi,T)}{\partial \chi^{2}},
\end{equation}
where
\begin{equation}\label{chi and a}
\chi =\frac{2}{3(1-\omega)} a^{\frac{3(1-\omega)}{2}},
\end{equation}
$\Psi$ is the wave function of the universe, $a$ is the scale factor, and $T$ is a parameter related to the perfect fluid, which plays the role of time through $dt=a^{3\omega}dT$. Note that for $\omega=1/3$, the parameter $T$ becomes equal to the conformal time $\eta$.

\begin{figure}[h]
    \centering
    \includegraphics[scale=0.9]{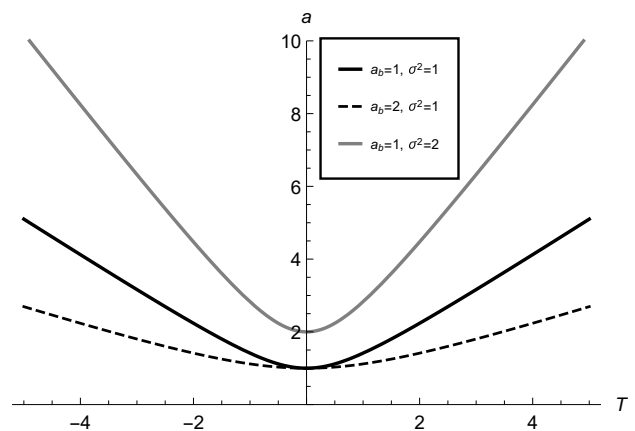}
    \caption{$a$ vs. $T$ for  $\omega=1/3$. The curves are obtained for some representative values of $a_b$ and $\sigma$.}
    \label{fig:bounce simetrico}
    \end{figure}

Imposing the Gaussian initial wave function at $T=0$
\begin{equation}\label{symmetric wave function}
\Psi_{0}(\chi)=\left( \frac{8}{\sigma^{2}\pi}\right)^{\frac{1}{4}} \exp \left( -\frac{\chi^{2}}{\sigma^{2}} \right),
\end{equation}
and implementing an unitary evolution, we obtain the wave function solution
\begin{eqnarray}
&& \nonumber \Psi(\chi,T)=\left[ \frac{8 \sigma^{2}}{\pi(\sigma^{4}+T^{2})} \right]^{\frac{1}{4}} \exp \left[-\frac{\sigma^{2}\chi^{2}}{\sigma^{4}+T^{2}}\right]\\
 &&\times \exp\left[ - i \left(\frac{T \chi^{2}}{\sigma^{4}+T^{2}}+\frac{1}{2} \arctan
 \left(\frac{\sigma^{2}}{T}\right)-\frac{\pi}{4}\right)\right].
\end{eqnarray}
Using the guidance equation of the dBB formalism,
\begin{equation}\label{trajectory equation in chi}
    \frac{d\chi}{dT}=-\frac{1}{2}\frac{\partial S}{\partial \chi} ,
\end{equation}
where $S$ is the phase of the wave function, 
the correspondent Bohmian trajectory for the scale factor $a$ reads
\begin{equation}\label{symmetric bounce in a}
a(T)=a_{b}\left[ 1+\left( \frac{T}{\sigma^{2}} \right)^{2} \right]^{\frac{1}{3(1-\omega)}},
\end{equation}
where $a_{b}$ is the value of the scale factor $a$ at the moment of the bounce $T=0$.
The expression (\ref{symmetric bounce in a}) describes a symmetric bounce, which corresponds to the classical solution for large values of $T$, and is plotted in Fig.~\ref{fig:bounce simetrico}. The value considered for the equation of state parameter, $\omega=1/3$, represents an universe filled with radiation fluid, as we expect for early times.

In order to relate the wave function parameters to observables, we obtain the Hubble parameter $H=\dot{a}/a$, where dot denotes derivative with respect to the physical cosmic time. For large values of T, the squared Hubble parameter reads
\begin{equation}\label{hubble2 simetrico}
H^{2}=\frac{a_{b}^{2}}{a^{4}\sigma^{4}}=H_0^2\Omega_{r0}\frac{a_0^4}{a^4},
    \end{equation}
where the subscript $_{0}$ in all quantities indicates their current values. We then identify the dimensionless density parameter for radiation today as
\begin{equation}\label{omegar0 simetrico}
        \Omega_{r0}=\frac{a_{b}^{2}}{a_{0}^{4}H_{0}^{2}\sigma^{4}}.
\end{equation}
Performing the change of variables given by $x_{b}=a_{0}/a_{b}$ and $\bar{\sigma}=\sigma \sqrt{a_{0}H_{0}}$, we obtain
\begin{equation}\label{sigma2 simetrico}
        \bar{\sigma}^{2}=\frac{1}{x_{b} \sqrt{\Omega_{r0}}}.
\end{equation}

The curvature scale at the bounce is given by
    \begin{equation}\label{Lb simetrico}
        L_{b}=\frac{1}{\sqrt{R}}\bigg. \bigg|_{T=0}=\frac{1}{x_{b}^{2}H_{0}\sqrt{6 \Omega_{r0}}},
    \end{equation}
where $R$ is the Ricci scalar.
It allows us to find lower and upper bounds to $x_{b}$ by requiring that the curvature scale at the bounce is some few orders of magnitude larger than the Planck scale (in order to
ensure that the Wheeler-DeWitt equation is a valid approximation of a more fundamental 
theory of quantum gravity \cite{Kiefer}), and 
smaller than the nucleosyntesis scale. As a result, we have
\begin{equation}\label{bounds symmetric}
    10^{11}\ll x_{b} < 10^{31}.
\end{equation}

\begin{figure}[h]
    \centering
    \includegraphics[scale=0.9]{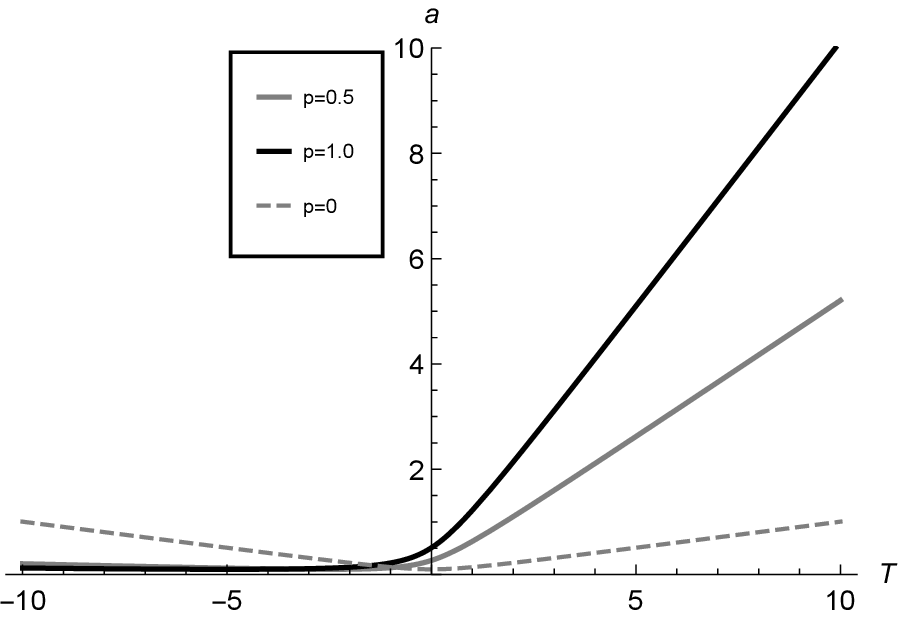}
    \caption{$a_{+}$ vs $T$ for $\sigma=1$, $a_{b}=1$, $\omega=\frac{1}{3}$.}
    \label{fig:bounce assimetrico nu 1}
    \end{figure}

\begin{figure}[h]
    \centering
    \includegraphics[scale=0.9]{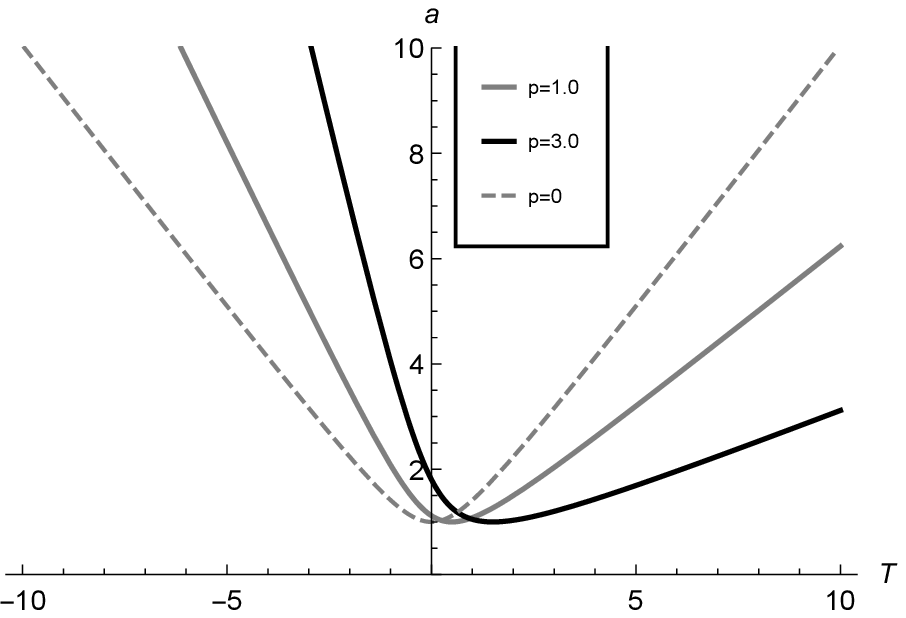}
    \caption{$a_{-}$ vs $T$ for $\sigma=1$, $a_{b}=1$, $\omega=\frac{1}{3}$.}
    \label{fig:bounce assimetrico nu 2}
    \end{figure}

We will also consider asymmetric solutions arising from the following initial wave function
\begin{equation}\label{non unitary initial wf}
   \Psi_{0}(\chi)=
    \exp \left( -\frac{\chi^2}{\sigma^2} \mp ip \chi    \right).
\end{equation}
Implementing a non-unitary evolution, which is detailed in \cite{Delgado:2020htr},
we obtain the resulting trajectory for the scale factor $a$, which is given by
\begin{eqnarray}\label{nu trajectory}
    \nonumber  a_{\pm}(T) &=& \Biggl \{ \Biggr. \pm \frac{3p(1-\omega)}{4}T+ a_{b}^{\frac{3(1-\omega)}{2} }
    \Biggl[ 1+ \left( \frac{T}{\sigma^{2}} \right) ^{2}\\
     &+&\left( \frac{3p(1-\omega)}{4} \right) ^{2}  \frac{(T^{2}+\sigma^{4})}{a_{b}^{3(1-\omega)}} \Biggr]^{\frac{1}{2}} \Biggl . \Biggr \} ^{\frac{2}{3(1-\omega)}},
\end{eqnarray}
where $a_{b}$ is the scale factor at the moment of the bounce $T_{b}=\mp \frac{p\sigma^{4}}{2 \chi_{b}}$ and $\chi_{b}$ is related to $a_{b}$ through Eq.~\eqref{chi and a}. The solutions $a_{+}$ and $a_{-}$ in Eq. (\ref{nu trajectory}) are plotted in Figs.~\ref{fig:bounce assimetrico nu 1} and \ref{fig:bounce assimetrico nu 2}, respectively, for $\omega=1/3$. The classical solution also
arises for large values of $T$ for both cases.

In this case, we perform the following change of variables:
 \begin{eqnarray}\label{tc assimetrico}
        x_{b}&=&\frac{a_{0}}{a_{b}},\\
        \bar{\sigma}&=&\sigma \sqrt{a_{0}H_{0}},\\
        \bar{p}&=&\frac{p}{a_{0}^{2}H_{0}},\\
        \bar{\eta}&=&\frac{T}{\sigma^2},\\
        y^2 &=& \frac{x_b\bar{p}{\bar{\sigma}}^2}{2},
\end{eqnarray}
which leads to a squared Hubble parameter in the expanding phase given by
\begin{equation}\label{Hpe}
            H^2=\frac{\biggl(\pm y^2+
        \sqrt{1+y^{4}}\biggr)^2 a_b^2 H_0^2 a_0^2}{{\bar{\sigma}}^4a^4}.
    \end{equation}
Thus the dimensionless density parameter of radiation today reads
\begin{equation}\label{omegar0 assimetrico}
    \Omega_{r0} =
    \frac{\left(\pm y^2+
        \sqrt{1+y^{4}}\right)^2}{{\bar{\sigma}}^4x_b^2} ,
    \end{equation}
while the wave function parameter $\bar{\sigma}$ reads
 \begin{equation}\label{sigma2 assimetrico}
\bar{\sigma}^{2}=
\left[
{ x_b^2 \Omega_{r0}(1\mp \bar{p}/\sqrt{\Omega_{r0}}})\right]^{-1/2}.
\end{equation}
As a consequence, for the solution $a_{+}$, the relation $\bar{p}<\sqrt{\Omega_{r0}}$ must be satisfied. As argued in \cite{Delgado:2020htr}, this solution is of special interest, once it can represent a bounce solution with an almost Minkowski contracting phase as $\bar{p}$ approaches $\sqrt{\Omega_{r0}}$.

For this asymmetric solution, the minimum curvature scale does not occur at the bounce, but at ${\bar{\eta}}_{\text{min}} = \mp \sqrt{\frac{\sqrt{1+y^4}-1}{2}}$. It reads
\begin{eqnarray}\label{Lm assimetrico}
\nonumber L_{\text{min}}&=&\frac{1}{\sqrt{R}}\bigg. \bigg|_{{\bar{\eta}}_{\text{min}}}\\ &=& \frac{R_{H0}\left(1+\sqrt{1\mp\frac{\bar{p}}{\sqrt{\Omega_{r0}}}}\right)^3}{8\sqrt{3\Omega_{r0}}x_{b}^{2}\left(1\mp\frac{\bar{p}}{\sqrt{\Omega_{r0}}}\right)^2\sqrt{\left(2\mp\frac{\bar{p}}{\sqrt{\Omega_{r0}}}\right)}}.
\end{eqnarray}
Note that, for the asymmetric solutions, the presence of extra parameter(s) related to asymmetry makes the physical bounds to be on $L_{\rm{min}}$ instead of $x_{b}$. They are given by
\begin{equation}
    10^{-58}\ll \frac{L_{\rm {min}}}{R_{H0}} < 10^{-20}.
\end{equation}

\begin{figure}[h]
    \centering
    \includegraphics[scale=1.0]{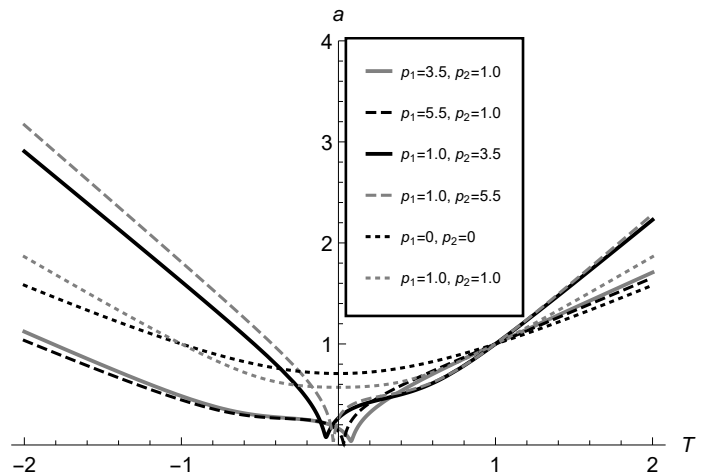}
    \caption{$a$ vs $T$ for $\sigma=1.0$, $a_{i}=1.0$, $T_{i}=1.0$ $\omega=\frac{1}{3}$.}
    \label{fig:bounce assimetrico u 1}
    \end{figure}

Another asymmetric solution is obtained by considering the following initial wave function
\begin{eqnarray}\label{wf0 unitary asymm}
\nonumber \Psi_{0}(\chi)&=& C \biggl[ \exp \left( -\frac{\chi^{2}}{\sigma^{2}}+i p_{1}^{2} \chi^{2} \right)\\
  &+& \exp \left( -\frac{\chi^{2}}{\sigma^{2}}-i p_{2}^{2} \chi^{2} \right) \biggr] ,
 \end{eqnarray}
where
\begin{eqnarray}
   \nonumber C&=&\frac{\sqrt{2}}{ \pi^{\frac{1}{4}}} \biggl \{   \left[ -i(p_{1}^{2}+p_{2}^{2})+\frac{2}{\sigma^{2}} \right]^{-\frac{1}{2}} \\
    &+&\left[ i(p_{1}^{2}+p_{2}^{2})+\frac{2}{\sigma^{2}} \right]^{-\frac{1}{2}} +\sqrt{2} \sigma \biggr\}^{-1/2}.
\end{eqnarray}
Implementing an unitary evolution
and following the dBB procedure, we obtain
a differential equation, which is shown in \cite{Delgado:2020htr} and can be solved numerically with initial condition $a_{i}=a(T_{i})$. The numerical solutions are plotted in Fig.~\ref{fig:bounce assimetrico u 1}. The classical limit arises for large values of $T$. This solution also encompasses multiple bounces. However, for our purpose in this work, we consider only the single bounce solutions.

For the limit where $p_{1}\sigma\ll1$ and $p_{2}\sigma\ll1$, it is possible to relate the wave function parameters to observables. The squared Hubble parameter reads
\begin{equation}\label{hubble2 assimetrico unitario}
        H^{2}=\frac{a_{i}^{2}}{a^{4}(T_{i}^{2}+\sigma^{4})},
    \end{equation}
allowing us to identify the dimensionless density parameter for radiation today as
\begin{equation}\label{omegar0 assimetrico unitario}
        \Omega_{r0}=\frac{a_{i}^{2}}{a_{0}^{4}H_{0}^{2}(T_{i}^{2}+\sigma^{4})}.
    \end{equation}
Replacing the initial values $T_i$ and $a_i=a_i(T_i)$ by $T_b$ and $a_{b}=a(T_{b})$, where $T_{b}=(p_{1}^{2}-p_{2}^{2})\sigma^{4}/2$, and performing the following change of variables
\begin{eqnarray}\label{tc assimetrico unitario p1 e p2}
    x_{b}&=&\frac{a_{0}}{a_{b}},\\
    \bar{\sigma}&=&\sigma \sqrt{a_{0}H_{0}},\\
    \bar{p}_{i}^{2}&=&\frac{p_{i}^{2}}{a_{0}H_{0}},
\end{eqnarray}
where $i=1,2$, we obtain the parameter $\bar{\sigma}^{2}$ as
    \begin{equation}\label{sigma2 assimetrico unitario p1 e p2}
    \bar{\sigma}^{2}=\left\{\frac{x_{b}^{2} \Omega_{r0}}{2}\left[1+\sqrt{1+\frac{(\bar{p}_{1}^{2}-\bar{p}_{2}^{2})^{2}}{x_{b}^{2}\Omega_{r0}}}\right]\right\}^{-1/2}.
\end{equation}
Note that $T_{b}$ appears in $\Omega_{r0}$ squared, thus $p_{1}\sigma$ and $p_{2}\sigma$ appear in fourth order. Disregarding these terms, Eq.~\eqref{sigma2 assimetrico unitario p1 e p2} reduces to Eq.~\eqref{sigma2 simetrico} of the symmetric case.

Since we are considering a limit in which the parameters related to asymmetry are small, the difference between the curvature scale at the bounce $L_{b}$ and the minimum curvature scale $L_{\text{min}}$ is not relevant. The expression for $L_{b}$ is given by Eq.~\eqref{Lb simetrico}, since $p_{1}\sigma$ and $p_{2}\sigma$ appear in fourth order, and are disregarded. As a consequence, the lower and upper bounds
\begin{equation}
    10^{-58}\ll \frac{L_{b}}{R_{H0}} < 10^{-20},
\end{equation}
reduce to Eq.~\eqref{bounds symmetric}.


\section{Baryogenesis with Curvature Coupling}
\label{secIV}

In recent decades, many suggestions have been made regarding the production of baryonic matter in the early Universe.  In this section, we explore the gravitational baryogenesis proposal of Ref.~\cite{Davoudiasl:2004gf}, ~implemented through a coupling term between the derivative of the Ricci curvature scalar, \(\partial_\mu R\), and the baryonic current, \(J^{\mu}\), in the cases of symmetric and asymmetric bounces of section III. 
We assume the coupling term to be a CP-violating interaction with the form, as proposed in Ref.~\cite{Davoudiasl:2004gf},
\begin{equation}
\frac{1}{M_*^2}\int d^4x\sqrt{-g}(\partial_\mu R)J^{\mu}, \label{eq1}
\end{equation}
where \(M_*\) is the cutoff energy scale of the effective theory. This term, in an expanding universe, also dynamically breaks CPT, and favors a net asymmetry towards the production of baryons over antibaryons. This happens because the \(J^{0} (=n_B)\) term has a different sign for matter versus antimatter, and as we will demonstrate here, it can be used to calculate the net asymmetry of matter and antimatter once the universe reaches a decoupling temperature for this effective theory.
Then, if we also assume that the characteristic  timescale \(\tau\) of the interaction runs faster than the expansion rate of the universe, that is, 
\begin{equation}
    \tau<<H^{-1},
\end{equation}
where \(H\) is the Hubble constant, we can use thermal equilibrium relations to calculate the baryon-to-entropy ratio. 
Identifying the term multiplying the baryon density \(n_B\) as its chemical potential $\mu_B$, namely
\begin{equation}
    \mu_B = \pm \frac{\dot{R}}{M_*^2},
\end{equation}
where the plus (minus) sign stands for particles (anti-particles), 
and using Eq.~\eqref{eqn: razão BF}, 
the baryon-to-entropy ratio at decoupling temperature reads
\begin{equation}\label{nbs}
\frac{n_B}{s}=\frac{15 g_B}{4\pi^{2}g_{*}}\frac{\dot{R}}{M_*^2 T}\Big|_{T=T_D},
\end{equation}
where \(s\) is the entropy, and \(T_D\) is the effective interaction decoupling temperature. Once the temperature drops below \(T_D\), the effective baryon production freezes, and the resulting asymmetry is then preserved.
In order to look for physical solutions, we must require this ratio to be the observed $n_B/s\approx9 \times 10^{-11}$, and then look to the parameter space in order to obtain 
the region satisfying it.

In order to calculate $n_B/s$, Eq.~\eqref{nbs}, the first step is to calculate $\dot{R}$.
The Ricci scalar is given by 
\begin{equation}\label{Rsym}
    R(\bar{\eta}) = \frac{6}{a_b^2 \sigma^4} \frac{A''(\bar{\eta})}{A^3(\bar{\eta})},
\end{equation}
where $\bar{\eta}\equiv \eta/\sigma^2$ is a dimensionless conformal time, $'\equiv d/d\bar{\eta}$,  and $A(\bar{\eta})\equiv a(\bar{\eta})/a_b$. 
The derivative of Ricci scalar, Eq.~\eqref{Rsym}, in cosmic time as a function of the conformal time then yields,
\begin{equation}
    \dot{R}(\bar{\eta})=\frac{6}{a_b^3 \sigma^6 }\frac{A(\bar{\eta})A^{'''}(\bar{\eta})-3 A'(\bar{\eta}) A''(\bar{\eta})}{ A^5(\bar{\eta})}.
\end{equation}
The next step is to obtain a relation of the type $\bar{\eta}=\bar{\eta}(T)$ far from the bounce, which is necessary in order to evaluate $n_B/s$ at $T=T_D$.
This can be accomplished by evaluating the relations $t=t(T)$ and $t=t(\bar{\eta})$, both far from the bounce. Combining theses relations, one obtains $\bar{\eta}=\bar{\eta}(T)$.
To obtain the first relation, we use the fact that, for reasonable values of \(T_D\), we expect the universe to still be dominated by radiation, and we can write the energy density as both
\begin{equation}\label{rho1}
\rho(t)=\frac{3 {M_p}^2}{32 \pi  t^2},
\end{equation}
and 
\begin{equation}\label{rho2}
\rho(T)=\frac{g_{*}\pi ^2 T^4}{30}, 
\end{equation}
where $M_p$ is the Planck mass.
From the equivalence between Eqs.~\eqref{rho1} and~\eqref{rho2}, we can get $t(T)$, which results in
\begin{equation}\label{tT}
t(T)=0.3\times g_*^{-1/2}\frac{M_P}{T^2}.
\end{equation}
In order to obtain $t=t(\bar{\eta})$, we need to specify the scale factor $A(\bar{\eta})$. Then, we can finally express the baryon-to-entropy ratio only as a function of $T$, and evaluate it at decoupling temperature $T_D$. From now on we will work with $g_{*}\approx 100$, and $g_{b}\approx 1$. 

We can already mention that Eq.~\eqref{nbs} will depend on the physical variables $x_b$, $T_D$ and $M_*$. The former is a red-shift variable at the bounce and it lies in the range $10^{15}\leq x_b\leq 10^{31}$, where the lower limit restricts the calculation to well before the start of nucleosynthesis, whereas the upper limit is due to the Planck scale. The decoupling temperature $T_D$ lies in the range $10\, \mathrm{TeV}\leq T_D\leq 10^{19}\,\mathrm{GeV}$, where the lower limit avoids observable effecs in LHC, and the upper limit is the Planck scale. Finally, for the cutoff energy scale $M_*$ we set the range $10^{-16} M_p \leq M_* \leq M_p$, for the same reasons as $T_D$. Therefore, we have
\begin{eqnarray}
\label{xB}10^{15}\leq &\,x_b\,&\leq 10^{31},\\
\label{TD}10^7\leq &\,\bar{T}_D\,&\leq 10^{22},\\
\label{Mstar} 10^{-16} \leq &\,\bar{M}_*\,& \leq 1,
\end{eqnarray}
where $\bar{T}_D=T_D/\mathrm{MeV}$ and $\bar{M}_*=M_*/M_\mathrm{p}$ are dimensionless quantities.

In the following we consider the cases of symmetric and unitary and non-unitary asymmetric bounces.

\subsection{Symmetric bounce}

For the symmetric case, we consider the bounce generated by the initial Gaussian wave function given by Eq.~\eqref{symmetric wave function}, which produces the scale factor given by Eq.~\eqref{symmetric bounce in a}. For a radiation dominated bounce ($\omega=1/3$), in conformal dimensionless time this reads:
\begin{equation}\label{asymrad}
a(\bar{\eta})=a_b\sqrt{1+\bar{\eta}^2},
\end{equation}
where $\bar{\eta}=\eta/\sigma^2$.
Far from the bounce, we can relate the cosmic time $t$ with $\bar{\eta}$ as:
\begin{equation}\label{teta}
t(\bar{\eta})=\frac{a_b \sigma^2\bar{\eta} ^2}{2},
\end{equation}
where, from Eq. \eqref{sigma2 simetrico}, $\sigma=\bar{\sigma}(a_{0}H_{0})^{-1/2}$ can be related to the Hubble radius $R_{H_0}=1/H_0$ and the radiation density today $\Omega_{r0}$ as 
\begin{equation}
     \sigma^2=\frac{R_{H_0}}{a_b x_b^2\sqrt{\Omega_{r0}}}.
\end{equation} 
Considering the values $H_0=1.22\times 10^{-61}M_p$, $M_p=1.22 \times 10^{22}\, \mathrm{MeV}$, from Eqs.~\eqref{tT} and~\eqref{teta} one obtains
\begin{equation}\label{etaT}
\bar{\eta}(\bar{T})=\frac{1.0\times 10^{-10}\, x_b}{\bar{T}}.
\end{equation}
For the the symmetric scale factor given by Eq.~\eqref{asymrad}, using Eq.~\eqref{etaT} in Eq.~\eqref{nbs}, the result for $n_B/s$ far from the bounce ($\bar{\eta}\gg1$) reads:
\begin{equation}\label{nBssym}
\frac{n_B}{s}=6.2\times10^{-86}
\frac{\bar{T}_D^7}{x_b^2\,\bar{M}_*^2},
\end{equation}
where we used $\Omega_{r0}=8\times10^{-5}$.
Finally, ${n_B}/{s}$ is given in terms of the parameters $x_b$, $\bar{T}_D$ and $\bar{M}_*$.

One must notice that Eq.~\eqref{etaT} was obtained through an expansion assuming that $\bar{\eta}\gg 1$, which imposes the condition~\footnote{To apply the condition $\bar{\eta}\gg 1$, we assumed $\bar{\eta}>10$ as sufficient because we expand terms $1+\bar{\eta}^2$. We use the same assumption for the asymmetric cases in the following sections.}:
\begin{equation}\label{xbT}
\frac{x_b}{\bar{T}}\gtrsim 1.0\times 10^{11}.
\end{equation}
Hence, this inequality must be considered together with conditions~\eqref{xB}~-~\eqref{Mstar} when we look for the regions of interest in the parameter space.

\begin{figure}[htb!]
	\includegraphics[scale=0.4]{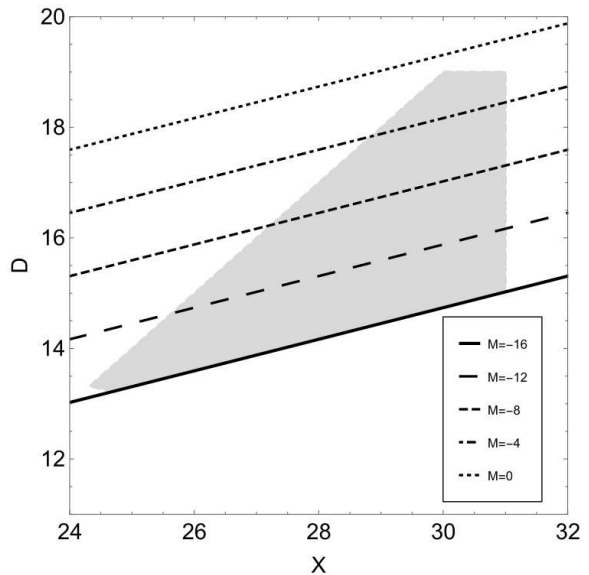}
	\caption{Parameter space of $x_B$, $T_D$, $M_*$ that that gives $n_B/s\approx9*10^{-11}$. These are parameterized by $X=\log (x_b)$, $D=\log (\bar{T}_D)$ and $M=\log(\bar{M}_*)$, respectively.}
	\label{sym}
\end{figure}

In Figure~\ref{sym}, we present a region plot for $\bar{T}_D\times x_B$ and lines for constant values of $\bar{M}_*$. The region of parameters that give $n_B/s\approx9*10^{-11}$ are the values of the gray region that are crossed by the constant $\bar{M}_*$ lines. 
In the following, we present analogous results for the asymmetric bounce cases, and we compare the results.

\subsection{Asymmetric bounce}

Let us now consider the case of asymmetric bounces, and look for the effects of asymmetry.
The calculations are similar to those of the symmetric case.
We also address the role of unitarity in the quantum evolution.
In the following, we choose two different cases, a non-unitary and a unitary asymmetric bounce.

\subsubsection{Non-Unitary Asymmetric Bounce}

For the non-unitary asymmetric case, we consider the bounce generated by the initial Gaussian wave function given by Eq.~\eqref{non unitary initial wf}, 
which produces the two classes of scale factors, given by Eq.~\eqref{nu trajectory}.
For a radiation dominated bounce in conformal dimensionless time, this reads:
\begin{eqnarray}\label{aasymnu}
a_{\pm}(\bar{\eta}) = a_b\left( \pm y^2\bar{\eta}
+ 
\sqrt{1+y^4}\sqrt{1+\bar{\eta}^2}
\right),
\end{eqnarray}
where $\bar{\eta}=T/\sigma^2$, $y^2=x_b\,\bar{p}\,\bar{\sigma}^2/2$, $\bar{\sigma}=\sigma \sqrt{a_0\,H_0}$ and $\bar{p}=p/(a_0^2\,H_0)$, assuming $p>0$, are dimensionless variables and the $\pm$ signals account for two possible bounce solutions. 
The bounce occurs at $\bar{\eta}_b=\mp y^2$.

As before, using the classical Friedmann equations in the expanding era far from the bounce, when quantum effects are negligible, 
we can obtain the relevant relations for our purpose.
The scale factor, Eq.~\eqref{aasymnu}, results:
\begin{eqnarray}\label{aasymnucl}
a(\bar{\eta}) = \frac{a_b\bar{\eta}}{\sqrt{1\mp \frac{\bar{p}}{\sqrt{\Omega_{r0}}}}}\quad , \quad (\eta\gg1).
\end{eqnarray}
The radiation density parameter $\Omega_{r0}$ is given by Eq. \eqref{omegar0 assimetrico},
which is used in order to express $\bar{\sigma}$ in the form of Eq. \eqref{sigma2 assimetrico}.
Also, due to the asymmetry of the scale factor, we can define a density parameter for the contracting phase, $\Omega_{cr}$ (the ratio between the radiation energy density and the critical density when the Hubble parameter at contraction has the same value as today), in terms of $\Omega_{r0}$, which reads:
\begin{eqnarray}\label{Omegacr}
\Omega_{cr}=\left(1\mp \frac{\bar{p}}{\sqrt{\Omega_{r0}}}\right)^2 \Omega_{r0}.
\end{eqnarray}
For simplicity, we define the following parameter:
\begin{eqnarray}\label{lambda}
\lambda = \sqrt{1\mp \frac{\bar{p}}{\sqrt{\Omega_{r0}}}},
\end{eqnarray}
observing the restriction $\bar{p}< \sqrt{\Omega_{r0}}$ for the $(-)$ ($a_+$) case.
In terms of this parameter, Eqs.~\eqref{aasymnucl}~-~\eqref{Omegacr} can be rewritten as:
\begin{eqnarray}
&&a(\bar{\eta})=\frac{a_b\bar{\eta}}{\lambda}
\quad,\quad
\Omega_{r0}=
\frac{1+\lambda ^4 \pm
(1-\lambda^4)}
{2\lambda^2 x_b^2\bar{\sigma}^4}   
,\nonumber\\
&&\bar{\sigma}^2=\frac{1}{x_b\sqrt{\Omega_ {r0}}
\lambda}
\quad,\quad
\Omega_{cr}=\lambda^4 \Omega_{r0},\nonumber
\end{eqnarray}
where $\lambda$ includes both $\pm$ cases. 
Using this unified description, we look for analytical constraints on the free parameters of the theory, which yield the observed baryon-to-entropy ratio.

Far from the bounce, we can relate the cosmic time $t$ with $\bar{\eta}$ as:
\begin{eqnarray}\label{tetaasym}
t(\bar{\eta})=
\frac{a_b \sigma^2 \bar{\eta}^2}{2  \lambda}.
\end{eqnarray}
From Eqs.~\eqref{tT} and~\eqref{tetaasym},
and using the parameters of the symmetric case, we obtain
\begin{eqnarray}\label{etaTasym}
\bar{\eta}(\bar{T})=
\frac{
1.0\times 10^{-10}x_b
\;\lambda
}
{\bar{T} }.
\end{eqnarray}
From the latter equation, our approximation demands we consider $\eta\gg1$, which results:
\begin{eqnarray}\label{xbTasym}
\frac{x_b}{\bar{T}}\gtrsim\frac{1.0\times 10^{11}}{\lambda}
\end{eqnarray}
As it was explained in section III, we have to use condition~\eqref{Lm assimetrico} on $L_\mathrm{min}=1/R(\bar{\eta}_\mathrm{{max}})$ directly. Defining the new parameter 
$\bar{L}=R_{H_0}/L_\mathrm{min}$, one gets
\begin{eqnarray}\label{Lmin}
10^{20}<\bar{L}<10^{58},   
\end{eqnarray}
which replaces~\eqref{xB}.
The lower limit is due to imposing $L_\mathrm{min}>10^3 l_\mathrm{p}$, where $l_\mathrm{p}$ is the Planck length, whereas the upper limit comes from the requirement that the bounce energy scale should be much bigger than the nucleosynthesis energy scale, $L_\mathrm{min}< L_\mathrm{nucl}$. 
From the definition of the Ricci scalar, Eq.~\eqref{Rsym}, we can write $\bar{L}$ as
\begin{eqnarray}
\label{Lmineq}
\bar{L}=
\frac{\sqrt{6\, \Omega_{r0}}\, x_b^2}{\mathcal{C}(\lambda)},
\end{eqnarray}
where 
\begin{eqnarray}
\label{C}
\mathcal{C}(\lambda)=\frac{(1+\lambda)^3}{4\sqrt{2}\lambda^2\sqrt{1+\lambda^2}}.
\end{eqnarray}
When $\mathcal{C}(1)=1$, $\bar{L}$ reduces to the symmetric result (see Ref.~\cite{Celani:2016cwm} for details.)
 
In terms of $\bar{L}$, $\bar{T}_D$, $\bar{M}_*$ and $\lambda$, the baryon-to-entropy ratio results
\begin{equation}\label{nBasym}
\frac{n_B}{s}=1.4\times 10^{-87}\, 
\frac{T_D^7}{\bar{M}_*^2\bar{L}}\, \mathcal{H}(\lambda),
\end{equation}
where
\begin{eqnarray}\label{H}
\mathcal{H}(\lambda) =\frac{2\sqrt{2}(1+\lambda^2)^{3/2}}{(1+\lambda)^3}= \frac{1+\lambda^2}{2\lambda^2\mathcal{C}(\lambda)}.
\end{eqnarray}
When $\mathcal{H}(1)=1$ and, using Eq.~\eqref{Lmineq} in Eq.~\eqref{nBasym}, we recover the symmetric result of Eq.~\eqref{nBssym}.

Now we look for the region in parameter space 
which gives $n_B/s\approx9\times 10^{-11}$.
Defining $S=\log \bar{L}$, $D=\log \bar{T}_D$ and $M=\log \bar{M}_*$, from Eq.~\eqref{nBasym}
one obtains:
\begin{eqnarray}\label{SDMlambda}  D(S,M,\lambda)=11+
0.14\left(S+2M -\log{\mathcal{H}(\lambda)}\right).
\end{eqnarray}
We need to apply the condition~\eqref{TD} to this equality and also consider conditions~\eqref{Mstar} and~\eqref{Lmin}.
In addition to these conditions, we also consider that $\eta\gg1$, since our results are valid far from the bounce. 
Imposing it to Eq.~\eqref{etaTasym}, and using Eq.~\eqref{Lmineq} to eliminate $x_b$, yields,
\begin{eqnarray}
S>2D +20.4 -0.434 \log {(\lambda^2\mathcal{C}(\lambda ))}.
\end{eqnarray}
Therefore, the complete set of conditions on the parameters are:
\begin{eqnarray}\label{full}
\label{TDasymNU}
&&7\leq \,D\,\leq 22,\\
\label{MstarasymNU}
&&-16 \leq \,M_*\, \leq 0,\\
\label{SasymNU}
&&20< \,S<\,58,\\
\label{ALL1asymNU}
&&S>2D +20.4 -0.434 \log {(\lambda^2\mathcal{C}(\lambda ))},\\
\label{ALL2asymNU}
&&D(S,M,\lambda)=11+
0.14\left(S+2M -\log{\mathcal{H}(\lambda)}\right).
\end{eqnarray}

However, we still want to use $x_b$ in order to compare to the results of the symmetric case. From the allowed values for $\bar{L}$, one can obtain the correspondent values for $x_b$. Writing $X=\log x_b$, from Eq.~\eqref{Lmineq} one obtains,
\begin{eqnarray}\label{XL}
S=2X-3.8-\log C(\lambda).   
\end{eqnarray}
This means that for each allowed values of $S$, there is a corresponding one for $X$.

From Eqs.~\eqref{ALL1asymNU},~\eqref{ALL2asymNU} and~~\eqref{XL},
one can note that asymmetry manifests itself due to the presence of the functions $\mathcal{C}(\lambda)$ and $\mathcal{H}(\lambda)$ on these conditions.
Inspecting the behavior of these functions, we can qualitatively realize that larger values of $\lambda$ imply smaller values of $x_b$, and greater values of $D$ and $\bar{M}$ as compared to the symmetric case, and the other way round for small $\lambda$, enlarging or shrinking the gray allowed region as a whole.

In order to confirm the previous analysis, 
we now consider some representative plots of these results in terms of $X$, $D$ and $M$ for some values of $\lambda$, for both ($\pm$) cases.

Let us first consider the (+) case, i.e., $a_+$ of Eq.~\eqref{aasymnu}.
In this case, there is a restriction
on $\bar{p}$ in Eq.~\eqref{lambda}, which sets the upper bound $\bar{p}<\sqrt{\Omega_{r0}}$. 
We expect that values of $\bar{p}$ close to this bound are the most relevant, since $\lambda\to0$ as $\bar{p}\to\sqrt{\Omega_{r0}}$, whereas smaller $\bar{p}$ tends to the symmetric case $\lambda\to1$. 

\begin{figure}[htb!]
	\includegraphics[scale=0.6]{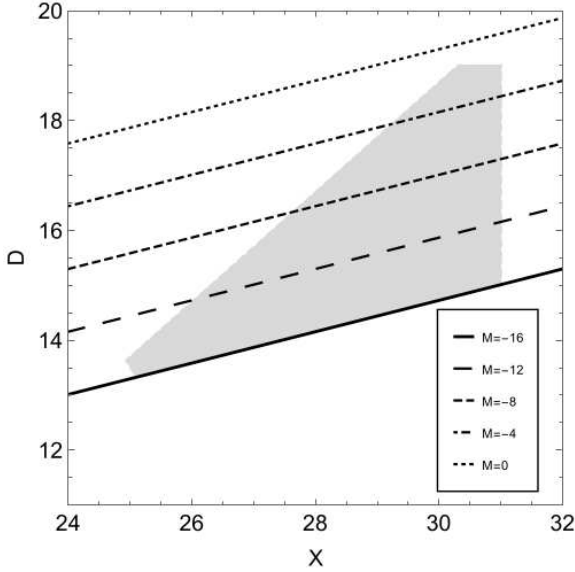}
	\caption{Parameter space of $x_B$, $T_D$, $M_*$ for $\bar{p}=0.50\sqrt{\Omega_{r0}}$ that that give $n_B/s\approx9*10^{-11}$. These are parameterized by $X=\log (x_B)$, $D=\log (\bar{T}_D)$ and $M=\log(\bar{M}_*)$, respectively, for scale factor $a_+$.}
	\label{asym0.5}
\end{figure}

\begin{figure}[htb!]
	\includegraphics[scale=0.6]{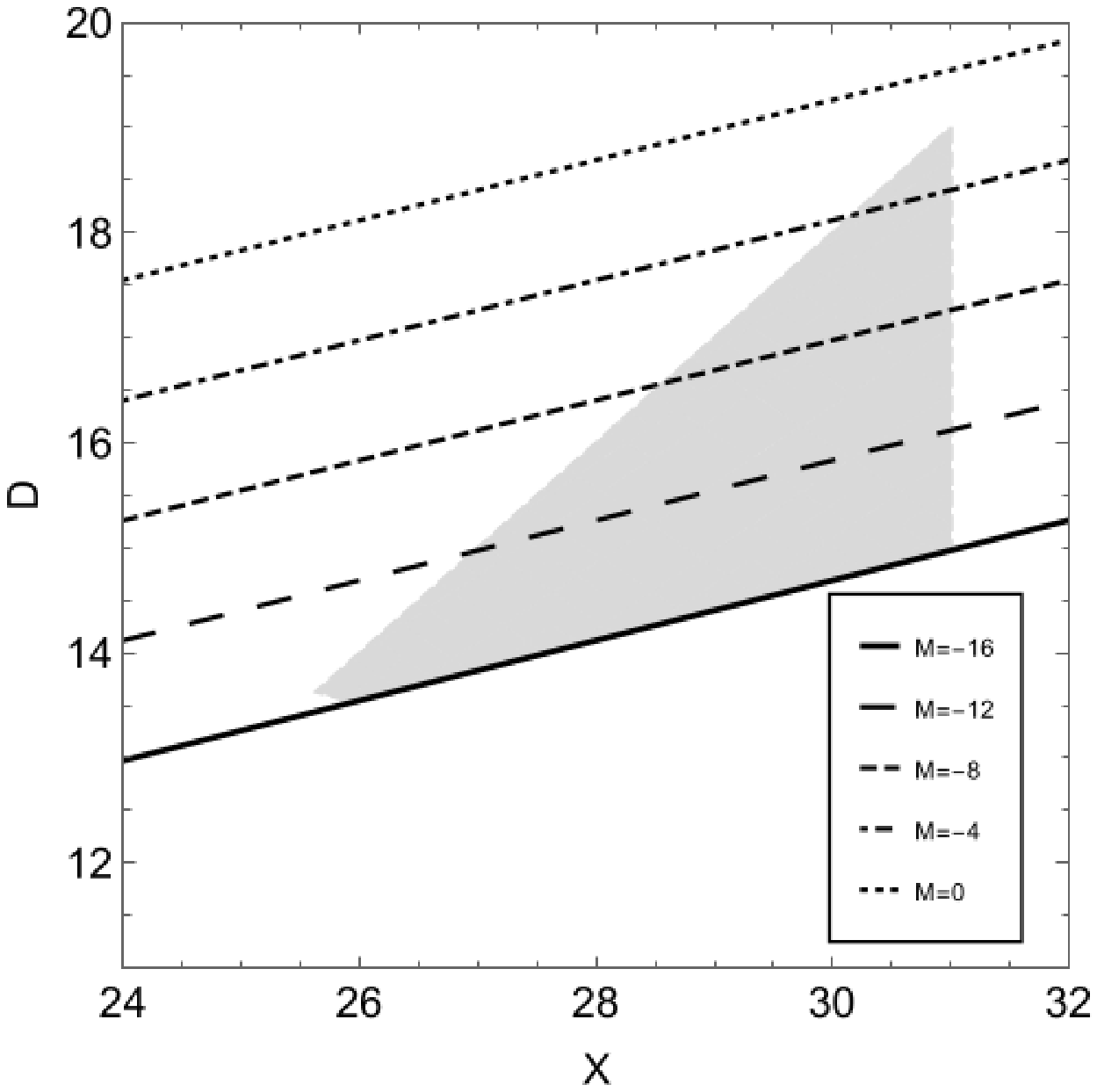}
	\caption{Parameter space of $x_B$, $T_D$, $M_*$ for $\bar{p}=0.90\sqrt{\Omega_{r0}}$ that that give $n_B/s\approx9*10^{-11}$. These are parameterized by $X=\log (x_B)$, $D=\log (\bar{T}_D)$ and $M=\log(\bar{M}_*)$, respectively, for scale factor $a_+$.}
	\label{asym0.9}
\end{figure}

\begin{figure}[htb!]
	\includegraphics[scale=0.6]{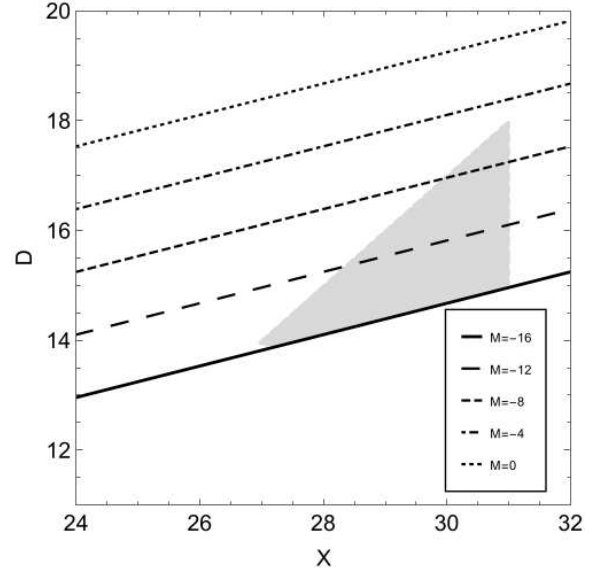}
	\caption{Parameter space of $x_B$, $T_D$, $M_*$ for $\bar{p}=0.99\sqrt{\Omega_{r0}}$ that that give $n_B/s\approx9*10^{-11}$. These are parameterized by $X=\log (x_B)$, $D=\log (\bar{T}_D)$ and $M=\log(\bar{M}_*)$, respectively, for scale factor $a_+$.}
	\label{asym0.99}
\end{figure}

In Figures~\ref{asym0.5},~\ref{asym0.9} and~\ref{asym0.99}, we note that as we increase $\bar{p}$ the gray ``triangle" is continuously smashed from the left, such that some smaller values of $X$ and higher values of $M$ are no longer allowed.
Therefore, the greater is the asymmetry (as $\lambda$ runs from $1$ to $0$), the smaller is the region of parameters compatible with the observed baryon-to-entropy ratio.

\begin{figure}[htb!]
	\includegraphics[scale=0.5]{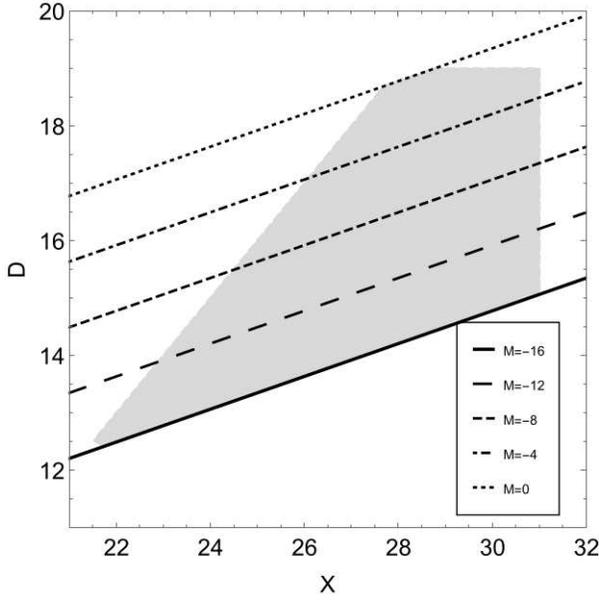}
	\caption{Parameter space of $x_B$, $T_D$, $M_*$ for $\bar{p}=10^2\sqrt{\Omega_{r0}}$ that that give $n_B/s\approx9*10^{-11}$. These are parameterized by $X=\log (x_B)$, $D=\log (\bar{T}_D)$ and $M=\log(\bar{M}_*)$, respectively, for scale factor $a_-$.}
	\label{asym100NEG}
\end{figure}

\begin{figure}[htb!]
	\includegraphics[scale=0.5]{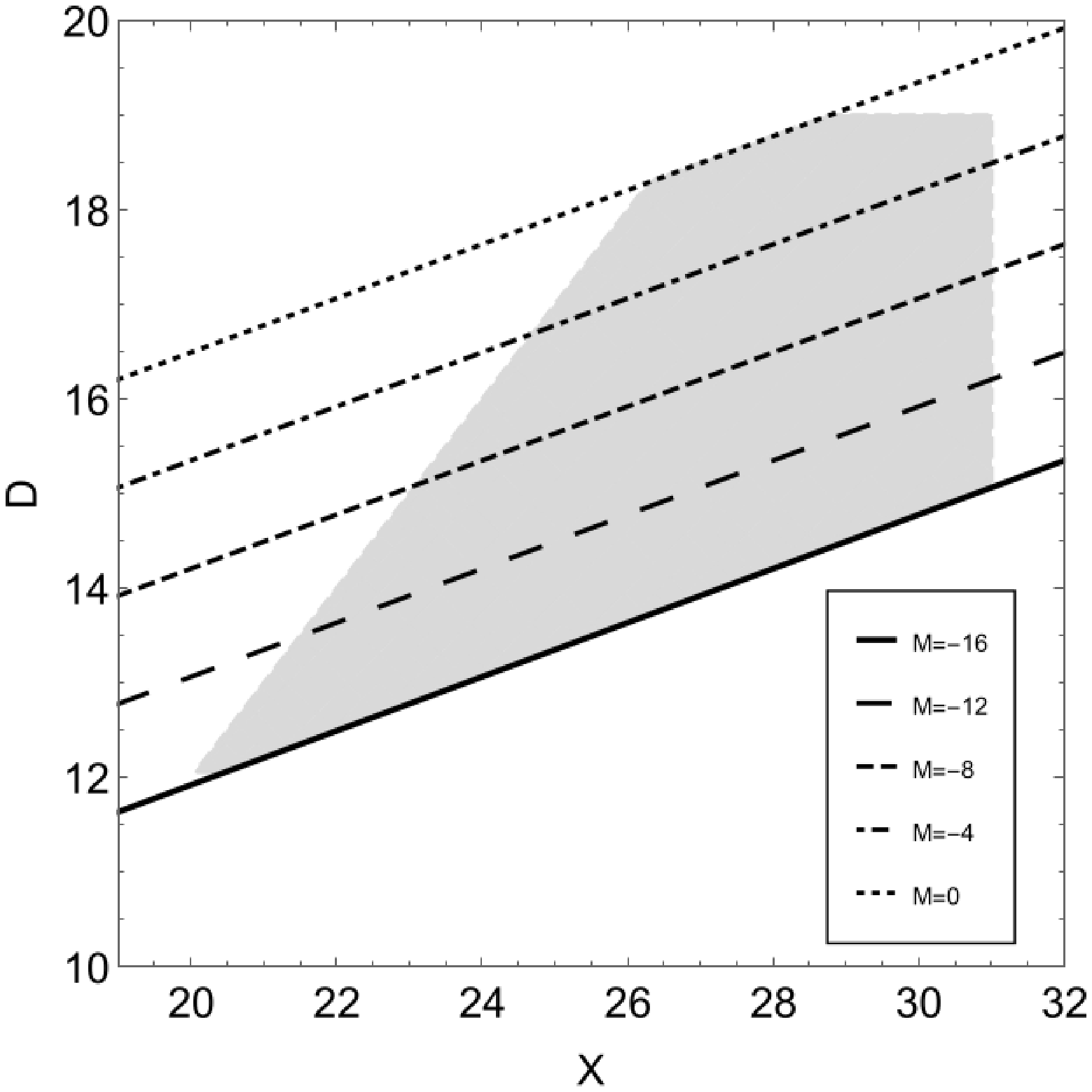}
	\caption{Parameter space of $x_B$, $T_D$, $M_*$ for $\bar{p}=10^3\sqrt{\Omega_{r0}}$ that that give $n_B/s\approx9*10^{-11}$. These are parameterized by $X=\log (x_B)$, $D=\log (\bar{T}_D)$ and $M=\log(\bar{M}_*)$, respectively, for scale factor $a_-$.}
	\label{asym1000NEG}
\end{figure}

\begin{figure}[htb!]
	\includegraphics[scale=0.5]{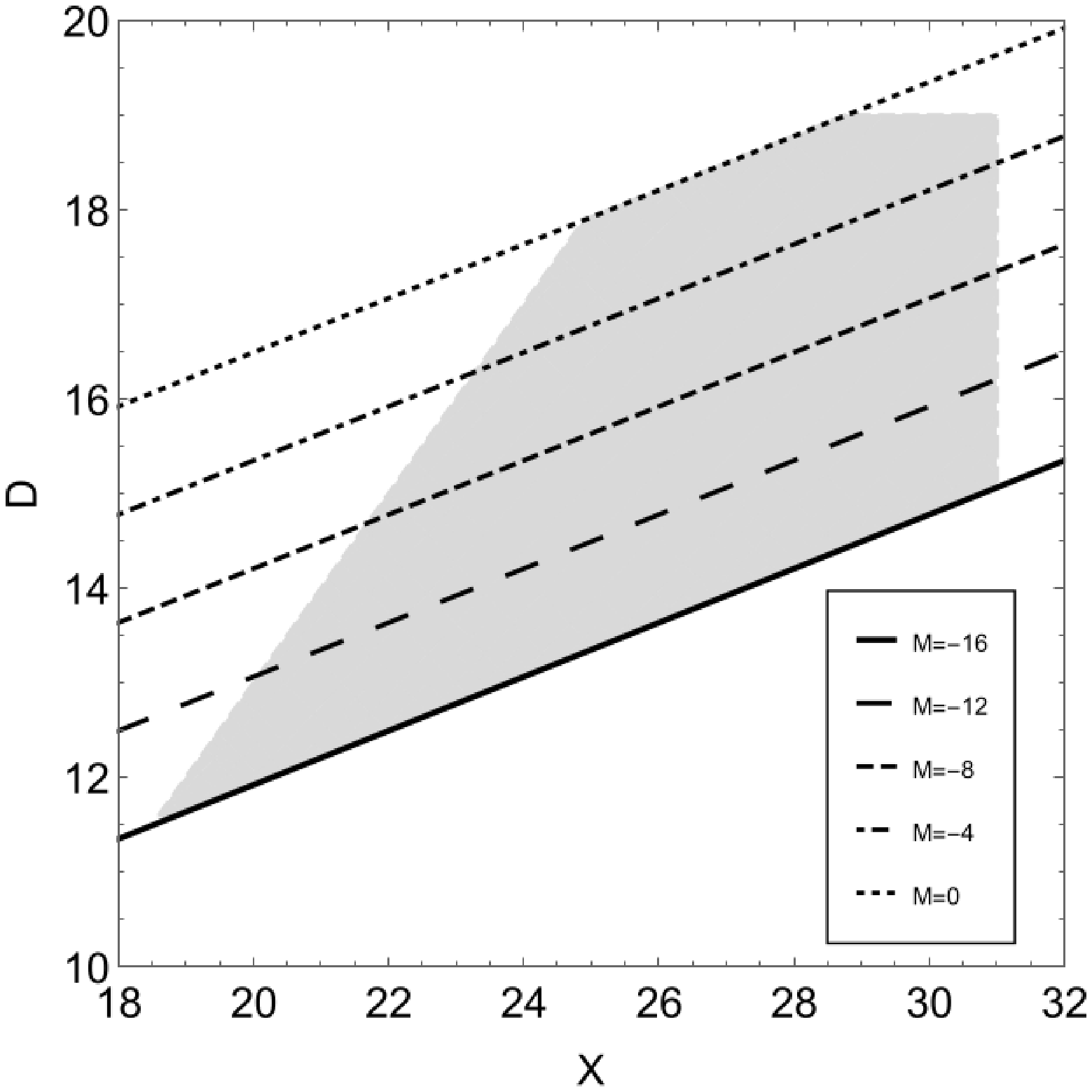}
	\caption{Parameter space of $x_B$, $T_D$, $M_*$ for $\bar{p}=10^4\sqrt{\Omega_{r0}}$ that that give $n_B/s\approx9*10^{-11}$. These are parameterized by $X=\log (x_B)$, $D=\log (\bar{T}_D)$ and $M=\log(\bar{M}_*)$, respectively, for scale factor $a_-$.}
	\label{asym10000NEG}
\end{figure}

\begin{figure}[htb!]
	\includegraphics[scale=0.5]{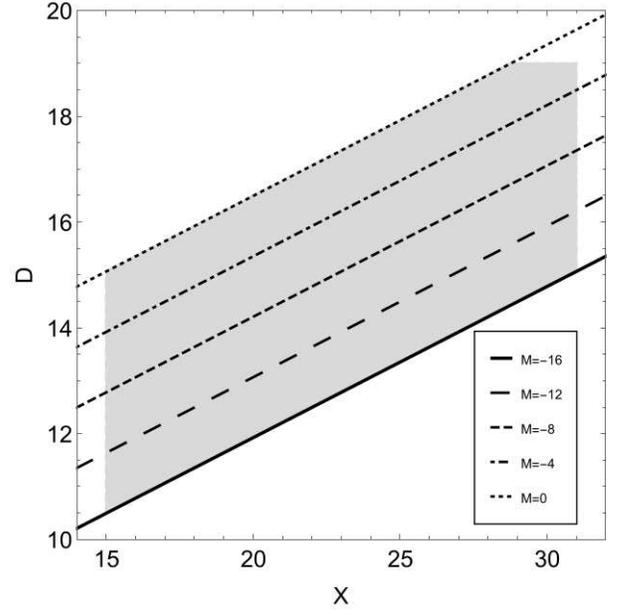}
	\caption{Parameter space of $x_B$, $T_D$, $M_*$ for $\bar{p}=10^{11}\sqrt{\Omega_{r0}}$ that that give $n_B/s\approx9*10^{-11}$. These are parameterized by $X=\log (x_B)$, $D=\log (\bar{T}_D)$ and $M=\log(\bar{M}_*)$, respectively, for scale factor $a_-$.}
	\label{asymP100000000000Omegar0NEG}
\end{figure}

In the (-) case, corresponding to the choice $a_-$ in Eq.~\eqref{aasymnu},
we note in Figs.~\ref{asym100NEG} 
-\ref{asymP100000000000Omegar0NEG}
that increasing $\bar{p}$ enlarge the allowed region in parameter space. 
The right side of this region does not change, whereas its left side grows to the left.  
As $\bar{p}$ increases, smaller $X$ (or $S$), and greater $D$ and $M$ are allowed 
in order to baryogenesis to occur with the right value.
For $\bar{p}\gtrsim 10^{11}\sqrt{\Omega_{r0}}$, the allowed region stabilizes: $X$, $M$ and $D$ reach their broader ranges.

In the following we consider another asymmetric bounce case, where there are two asymmetry parameters instead of one.

\subsubsection{Unitary Asymmetric Bounce}

We now consider the unitary asymmetric bounce given by the initial wave function of Eq.~\eqref{wf0 unitary asymm}, from which we can obtain a solution for the trajectory of the scale factor $a(T)$. 
The differential equation for $a(T)$ is rather involved (see Eq.~(54) of Ref.~\cite{Delgado:2020htr}), so that we perform some approximations on $a(T)$.
Since the relation between $\bar{\sigma}$ and the dimensionless energy density parameter of radiation today $\Omega_{r0}$, given by Eq.~\eqref{sigma2 assimetrico unitario p1 e p2}, is obtained in the limit $p_{1}\sigma\ll 1$ and $p_{2}\sigma\ll 1$, the analysis of baryogenesis with curvature coupling in this section also refers to this expansion. In this limit, it is possible to obtain an analytical expression for the scale factor by considering the terms $p_{1}\sigma$, $p_{2}\sigma$ and $\sqrt{p_{1}\,p_{2}}\,\sigma$ only up to second order. It reads
\begin{equation}\label{aexpunitary}
a(\bar{\eta})=
a_{b}
\left[ 1-\frac{(p_{1}^{2}-p_{2}^{2})\sigma^{2}\bar{\eta}}{2(1+\bar{\eta}^{2})}\right]
\sqrt{1+\bar{\eta}^{2}}.
\end{equation}

From now on we apply the transformation of variables $x_{b}=a_{0}/a_{b}$, $\bar{\sigma}=\sigma\sqrt{a_{0}H_{0}}$ and $\bar{p}_{i}^{2}=p_{i}^{2}/a_{0}H_{0}$, $i=1,2$.

Far from the bounce, the scale factor reads:
\begin{equation}\label{scaleasympt}
a(\bar{\eta})=a_b \bar{\eta},
\end{equation}
such that the cosmic time $t$ in terms of $\bar{\eta}$ gives
\begin{equation}\label{tfceta asymuni}
t(\bar{\eta})=\frac{a_b\sigma^{2}{\bar{\eta}}^{2}}{2},
\end{equation}
which are both identical to the symmetric case.
As we did for the previous cases, matching Eqs.~\eqref{rho1} and~\eqref{rho2}, we are able to find a relation between $\bar{\eta}$ and the temperature $\bar{T}$. Using Eq.~\eqref{tfceta asymuni}, we obtain
\begin{equation}\label{etafuncT}
\bar{\eta}(\bar{T})=
\frac{
1.0\times 10^{-10}x_b}{\bar{T} },
\end{equation}
which is also identical to the symmetric bounce.
As mentioned before, disregarding the terms $(\bar{p}_{1}\bar{\sigma})^{4}$, $(\bar{p}_{2}\bar{\sigma})^{4}$ and $(\bar{p}_{1}\bar{p}_{2})^{2}\bar{\sigma}^{4}$, we obtain that $\bar{\sigma}$, given by Eq.~\eqref{sigma2 assimetrico unitario p1 e p2}, also reduces to Eq.~\eqref{sigma2 simetrico} of the symmetric case.

As we did for the other bounce solutions,
using Eqs.~\eqref{aexpunitary} and~\eqref{etafuncT} and the constants defined in the symmetric case, we obtain $n_{B}/s$ equal to symmetric result, Eq.~\eqref{nBssym}, up to seventh order, plus an extra term in the eighth order: 
\begin{eqnarray}
\frac{n_{B}}{s}&=&
\frac{n_{B}}{s}\bigg|_{\mathrm{sym}}
+2.8\times 10^{-73} \frac{\bar{T}_{D}^{8}}{\bar{M}_{\star}^2 x_{b}^{4}}|\bar{p_{1}}^{2}-\bar{p_{2}}^{2}|.
\end{eqnarray}
Note that the maximum value of the new term corresponds to the maximum value of the difference $|\bar{p_{1}}^{2}-\bar{p_{2}}^{2}|$.

In order to assure $\bar{\eta} \geqslant 1$ ($\bar{\eta}\gtrsim 10$), we find the same condition obtained for the symmetric bounce:
\begin{eqnarray}\label{ineqxTD}
\frac{x_{b}}{\bar{T}}\gtrsim 1.0\times 10^{11}.
\end{eqnarray}

Finally, considering reasonable values of $x_{b}$, $T_{D}$ and $M_{\star}$, given by conditions~\eqref{xB}~-~ \eqref{Mstar},
we obtain the region of parameters that allows $n_{B}/s=9\times10^{-11}$.
Considering $n_{B}/s$ up to eighth order, we obtain the same plot of the symmetric case, given by Figure~\ref{sym}, for any value of the parameters $\bar{p_{1}}$ and $\bar{p_{2}}$ that satisfy $\bar{p_{1}}\bar{\sigma}\ll 1$ and $\bar{p_{2}}\bar{\sigma}\ll 1$. 
This means that the eighth order does not bring new possibilities of parameters allowed.
Hence, the gravitational baryogenesis of the unitary asymmetric bounce in this limit is equal to the gravitational baryogenesis of the symmetric case.

\section{Baryogenesis with Scalar Coupling}
\label{secV}

The action for a canonical scalar field $\phi$ in a curved space-time with metric $g_{\mu \nu}$ reads
 \begin{equation}\label{eqn:Ação}
     S = \int d^{4}x\sqrt{-g}\left[\frac{1}{2}(\partial_{\mu} \phi)^{2} - V(\phi) \right].
 \end{equation}\label{eqn: potencial} 
As discussed in Refs.~\cite{Cohen:1987vi,DeSimone:2016ofp}, spontaneous baryogenesis can be driven by the coupling of the baryonic current with the derivative of the scalar field ${\partial}_{\mu}\phi$ through
\begin{equation}
\frac{1}{M_*}\int d^4x\sqrt{-g}(\partial_\mu \phi)J^{\mu}. \label{eq1}
\end{equation}
The chemical potential in this case is given by
\begin{equation}\label{ref:potencial quimico}
\mu_B = \frac{\dot\phi}{M_{*}},
 \end{equation}
where $M_*$ is the energy scale of the coupling. Again, in a dynamical universe where $\phi$ evolves in time, CPT-invariance is broken, as in the curvature coupling discussed in the previous section, and baryons can be created even in thermal equilibrium. 

In this framework, the bounce background dynamics is given in Ref.~\cite{Bacalhau:2017hja}, where a scalar field with exponential potential drives the bounce as a stiff matter fluid, behaves as a dust fluid in the asymptotic past and future (which guarantees an almost scale invariant spectrum of scalar perturbations, see Ref.~\cite{Bacalhau:2017hja}), and also presents a transient dark energy-type behavior occurring only in the future of the expanding phase. This bounce is asymmetric because the transient dark energy epoch occurs only in the expanding phase, not in the contracting phase, avoiding problems 
related to the imposition of vacuum state initial conditions in the contracting phase if dark energy is present there, and overproduction of gravitational waves, which
are typical in bouncing models containing a canonical scalar field.


The pressure and energy density associated with $\phi$ are, respectively,
\begin{eqnarray}
\label{eqn:pressão} P_\phi&=&\frac{1}{2}\dot\phi^2 - V(\phi),
\\
\label{eqn:densidade} \rho_\phi&=&\frac{1}{2}\dot\phi^2 + V(\phi),
\end{eqnarray}
and the potential reads
\begin{equation}
\label{def_pot}
V(\phi) = V_0 e^{-\lambda \kappa \phi},
\end{equation}
where the constant $V_0$ has units of mass to the fourth power, and $\lambda$ is dimensionless,
chosen to satisfy $\lambda \approx \sqrt{3}$ in order to get an almost scale invariant spectrum of scalar perturbations.
The background dynamics can be made simpler through a choice of
dimensionless variables, 
\begin{equation}\label{mud1}
x = \frac{\sqrt{8\pi} \dot{\phi}}{\sqrt{6}M_P H}, \qquad
y = \frac{\sqrt{8\pi V}}{\sqrt{3}M_P H}.
\end{equation}
In these new variables, the Friedmann constraint and the effective equation of state parameter, $w=P/\rho$, read,
\begin{equation}
x^2 + y^2 = 1, \qquad  w = 2x^2-1. \label{x_y_con}
\end{equation}
The above definitions lead to the planar
system:
\begin{align}
\frac{\dd x}{\dd \alpha} &= - 3 x (1-x^2)+ \lambda \sqrt{\frac{3}{2}}y^2, \label{sist1} \\
\frac{\dd y}{\dd \alpha} &= x y \left(3x-\lambda \sqrt{\frac{3}{2}}\right), \label{sist2}
\end{align}
where $\alpha \equiv \ln (a)$. This system is supplemented by the
equations
\begin{equation}\label{eq:sup}
\dot{\alpha} = H, \qquad \dot{H} = -3H^2x^2.
\end{equation}
In the expanding phase, the variable $x$ is close to $1$ near the bounce (the kinetic
part dominates, hence $w\approx 1$, as stiff matter), and decrease to $0$ (an effective $w\approx -1$, as dark energy),
passing in between through $x=\sqrt{2/3}$, or $w\approx 1/3$, where it is connected with the standard cosmological
evolution before nucleosynthesis. Baryon production should terminate in this epoch. Afterwards the baryon number "freezes" in the current value of $n_B/s \approx  9\times 10^{-11}$~\cite{DeSimone:2016ofp}. 
Hence, the decoupling takes place when 
\begin{equation} 
P_\phi\approx\frac{1}{3}\rho_\phi,
\end{equation}
which, from Eqs.~\eqref{eqn:pressão} and~\eqref{eqn:densidade}, yields
\begin{equation}\label{eqn:densidade em função do campo}
        \rho_\phi\approx\frac{3}{4}\dot\phi^2.
\end{equation}
The radiation density, in turn, is given by the Stefan Boltzmann's law,
\begin{equation}\label{eqn:Stefan Bolt}
    \rho_r = \frac{\pi^2g_{*}}{30} T^4,
\end{equation}
which, together with Eq.~\eqref{eqn:densidade em função do campo}, implies that
\begin{equation}\label{eqn:derivada de phi em função de T}
\dot\phi\approx\sqrt{\frac{2\pi^2 g_{*}  }{45}}T^2.
\end{equation}
Hence, the chemical potential (\ref{ref:potencial quimico}) reads
\begin{equation}\label{eqn:novo potencial quimico}
\mu_B\approx\sqrt{\frac{2\pi^2 g_{*}  }{45}}\frac{T^2}{M_*}.
\end{equation}
The baryon-photon (entropy) ratio, given in terms of the decoupling temperature $ T_D $ 
and the coupling parameter $M_*$, now reads
\begin{equation}\label{eqn:novo xB}
\frac{n_B}{s}\approx\sqrt{\frac{5}{8\pi^2g_{*}}}\frac{T_D}{M_{*}}.
\end{equation}
Hence, we obtain the free parameter condition
\begin{equation}\label{parameters-scalar}
\frac{M_{*}}{T_D} \approx 2.8\times 10^{8}.
\end{equation}
For $M_* < M_P$ and $T_D > 10$\,TeV, one gets 
\begin{eqnarray}
\label{conditions-phi}
&&\,\,10^4\,\rm{GeV} < T_D < 10^{11}\,\rm{GeV}, \\
&&10^{12}\,\rm{GeV} < M_* < M_P.    
\end{eqnarray}

\section{Conclusion}
\label{conclusions}

In this paper, we studied cosmological baryogenesis in the context of gravitational and spontaneous baryogenesis. In both approaches, there is a new coupling of the baryon current with the gradient of either the Ricci scalar, or a scalar field, respectively. With this type of coupling, due to the absence of a time-like Killing vector in a dynamical universe, CPT-invariance, is broken and a net baryon number can emerge, even in thermal equilibrium. 

Making use of these proposals, we analyzed baryon production in background bouncing models, extending investigations already realized in the context of loop quantum cosmology~\cite{Odintsov:2016apy} to bouncing models coming from Wheeler-DeWitt quantum cosmology in the framework of the dBB quantum theory~\cite{Delgado:2020htr,Bacalhau:2017hja}. We investigated many possible bouncing solutions, symmetric and asymmetric around the bounce. The free parameters are the energy scale of the coupling, the curvature scale at the bounce, and the decoupling temperature.

In the case of gravitational baryogenesis, the results for the symmetric bounce shows that a broad region of physical parameters fulfil the  observed value of the baryon-to-entropy ratio. It is allowed by almost all possible values of coupling energy scales, with preference for the lower ones, a region of seven orders of magnitude for the curvature scale at the bounce, with preference for the deeper ones, and a narrower interval for possible decoupling temperatures (four orders of magnitude), with preference for the lower ones. 

Note that the bounce background is not a necessary condition for a net baryon number within gravitational and spontaneous baryogenesis, but it allows a larger range of parameters. In the case of gravitational baryogenesis, the decoupling temperature must be attained when $\dot{R}$ is not negligible. In classical cosmology, dominated by fields satisfying an effective equation of state $p=w\rho$, one has $\dot{R}=-24\pi G H(1-3w)(1+w)\rho$, hence it is negligible in a radiation dominated phase, or even during inflation. Of course, in the primordial Universe $w$ is not exactly $1/3$, but it is quite close, and hence the range of parameters necessary to yield sufficient baryogenesis is more constrained. In a quantum bounce, one has $\dot{R}=-24\pi G H_0[(1-3w)(1+w)\rho + Q]$, where $Q$ denotes quantum corrections (in the symmetric case of sub-section IV-A, one has $Q=(\rho_0 x^6)/x_b^{3(1-w)}$, where $x=a_0/a$ and $\rho_0$ is the energy density of the background fluid today), one has a larger range of possibilities, as we have seen in the paper. Any other modifications of the standard cosmological background, like in Loop Quantum Cosmology~\cite{Odintsov:2016apy}, with ghost condensates~\cite{ghost}, or Gauss-Bonnet corrections~\cite{GB}, tends to yield similar effects.

Setting the symmetric case as the basis for comparison, in the case of asymmetric bounces the region of allowed parameters is enlarged in the case where the contracting phase has more radiation energy than its expanding phase value, and it gets shrunken when the contracting phase has less radiation energy than its expansion value. In the limiting case of an almost empty contraction, no baryon asymmetry is obtained, even in the presence of the observed radiation energy density in the expanding phase.

From the discussion above, one can see that the net amount of baryons in the expanding phase does depend on what happens in the contracting phase, even keeping the same asymptotic amount of radiation in the expanding phase. This is because the whole dynamics of the background model emerges from the wave function of the universe itself. Modifications of the properties of the contracting phase are obtained by different choices of its parameters, which change the values of $Q$ present in $\dot{R}$, hence altering the amount of baryons in the expanding phase. A pure expanding universe with the same evolution of the scale factor in the expansion branch of our work would lead to the same result, but it is hard to imagine a physical framework in which such specific expanding phase might emerge without a preceding bounce. Note that baryogenesis also happens in the contracting phase, as long as interactions where baryon number is not conserved are effective, and $\dot{R}$, or $\dot{\phi}$, is not negligible, but the relevant quantity is the net baryon-antibaryon asymmetry at the decoupling temperature in the expanding phase.

For spontaneous baryogenesis, driven by a time-dependent scalar field, the background was the one of reference~\cite{Bacalhau:2017hja}, which is necessarily asymmetric. In this case, the baryon asymmetry depends only on the energy scale of the coupling, and the decoupling temperature: the background bouncing parameters mildly affect the classical limit of the model, and they are constrained by the amplitude and spectra of scalar cosmological perturbations coming from Cosmic Background Radiation observations. The allowed coupling energy scales cannot be very far from the Planck energy, whereas the allowed decoupling temperatures cannot be much bigger than its lower values fixed by observations.

Concluding, under the framework of gravitational and spontaneous baryogenesis, bouncing models can naturally yield the observed baryon asymmetry in the Universe, excepting the limiting case in which the contracting phase is almost empty, even with huge particle production at the bounce. It should be nice to study spontaneous baryogenesis in other bouncing scalar field models~\cite{ghost}, as well as for quantum bounces originated from other approaches~\cite{Peter-Sandro}. 

\begin{acknowledgments}
N.P.-N. would like to thank CNPq Grant No. PQ-IB
309073/2017-0 of Brazil for financial support. P.C.M.D. would like to thank CAPES Grant No. 88882.332430/2019-01 for financial support. T.M.C.S. would like to thank CAPES Grant No. 88882.332434/2019-01 for financial support.
\end{acknowledgments}

\appendix


\end{document}